\documentclass[a4paper,11pt]{article}
\usepackage{latexsym}
\newcommand{\BQED}{\hfill \hbox{\rule{8pt}{8pt}}}
\catcode`\@=11
\def\newexample#1{\@ifnextchar[{\@oexm{#1}}{\@nexm{#1}}}

\def\@nexm#1#2{%
\@ifnextchar[{\@xnexm{#1}{#2}}{\@ynexm{#1}{#2}}}

\def\@xnexm#1#2[#3]{\expandafter\@ifdefinable\csname #1\endcsname
{\@definecounter{#1}\@addtoreset{#1}{#3}%
\expandafter\xdef\csname the#1\endcsname{\expandafter\noexpand
  \csname the#3\endcsname \@exmcountersep \@exmcounter{#1}}%
\global\@namedef{#1}{\@exm{#1}{#2}}\global\@namedef{end#1}{\@endexample}}}

\def\@ynexm#1#2{\expandafter\@ifdefinable\csname #1\endcsname
{\@definecounter{#1}%
\expandafter\xdef\csname the#1\endcsname{\@exmcounter{#1}}%
\global\@namedef{#1}{\@exm{#1}{#2}}\global\@namedef{end#1}{\@endexample}}}

\def\@oexm#1[#2]#3{\expandafter\@ifdefinable\csname #1\endcsname
  {\global\@namedef{the#1}{\@nameuse{the#2}}%
\global\@namedef{#1}{\@exm{#2}{#3}}%
\global\@namedef{end#1}{\@endexample}}}

\def\@exm#1#2{\refstepcounter
    {#1}\@ifnextchar[{\@yexm{#1}{#2}}{\@xexm{#1}{#2}}}

\def\@xexm#1#2{\@beginexample{#2}{\csname the#1\endcsname}\ignorespaces}
\def\@yexm#1#2[#3]{\@opargbeginexample{#2}{\csname
       the#1\endcsname}{#3}\ignorespaces}

\def\@exmcounter#1{\noexpand\arabic{#1}}
\def\@exmcountersep{.}
\def\@beginexample#1#2{\trivlist \item[\hskip 
\labelsep{\bf #1\ #2:}]}
\def\@opargbeginexample#1#2#3{\trivlist
      \item[\hskip \labelsep{\bf #1\ #2\ }#3{\bf :}]}
\def\@endexample{\endtrivlist}

\catcode`\@=12

\newtheorem{lemma}{{\bf Lemma}}[section]
\newtheorem{thm}{{\bf Theorem}}[section]

\newtheorem{claim}{{\bf Claim}}[section]

\newtheorem{observation}{{\bf Observation}}[section]
\newcommand{\msc}[1]{\mbox{{\sc #1}}}
\catcode`\@=11

\def\@xnthm#1#2[#3]{\expandafter\@ifdefinable\csname #1\endcsname
{\@definecounter{#1}\@addtoreset{#1}{#3}%
\expandafter\xdef\csname the#1\endcsname{\expandafter\noexpand
\bf \csname the#3\endcsname \@thmcountersep \@thmcounter{#1}}%
\global\@namedef{#1}{\@thm{#1}{#2}}\global\@namedef{end#1}{\@endtheorem}}}

\def\@ynthm#1#2{\expandafter\@ifdefinable\csname #1\endcsname
{\@definecounter{#1}%
\expandafter\xdef\csname the#1\endcsname{\bf \@thmcounter{#1}}%
\global\@namedef{#1}{\@thm{#1}{#2}}\global\@namedef{end#1}{\@endtheorem}}}

\def\@begintheorem#1#2{\trivlist \item[\hskip 
\labelsep{\bf #1~#2:}]\sl}
\def\@opargbegintheorem#1#2#3{\trivlist
      \item[\hskip \labelsep{\bf #1~#2}~#3{\bf :}]\sl}

\catcode`\@=12

\newbox\rubisita
\newbox\rubiue
\newdimen\rubiw
\def\rubi#1#2{{\setbox\rubisita=\hbox{#1}\setbox\rubiue=\hbox{\tiny #2}%
\ifdim \wd\rubisita>\wd\rubiue\rubiw=\wd\rubisita\else\rubiw=\wd\rubiue\fi%
\kanjiskip=0pt plus1fil%
\setbox\rubisita=\hbox to \rubiw{\hfil#1\hfil}%
\setbox\rubiue=\hbox to \rubiw{\tiny\hfil#2\hfil}%
\vbox{\offinterlineskip\box\rubiue\break\box\rubisita}}}
\begin{document}
\begin{center}
{\Large {\bf  Buffer Management of Multi-Queue QoS Switches}}\\
{\Large {\bf  with Class Segregation}}\bigskip\\
\begin{tabular}{ccc}
{\sc Toshiya Itoh} & & {\sc Seiji Yoshimoto}\\
{\sf titoh@ip.titech.ac.jp} & & {\sf yoshimoto.s.aa@m.titech.ac.jp}\\
Imaging Science \& Engineering Laboratory & & 
Department of Information Processing\\
Tokyo Institute of Technology &  & 
Tokyo Institute of Technology\\
\end{tabular}
\end{center}\medskip
{\bf Abstract:} In this paper, we focus on buffer management of 
multi-queue QoS switches 
in~which~packets of different values are segregated in different queues. 
Our model consists of $m$ queues~and~$m$~packet~values $0<v_{1}<v_{2}<\cdots<v_{m}$.  Recently, 
Al-Bawani and Souza [IPL 113(4), pp.145-150,~2013] 
presented an online algorithm {\sc greedy} for 
buffer management of multi-queue QoS switches with~class~segregation 
and showed that if $m$ queues have the same size, then 
the competitive ratio of {\sc greedy}~is~$1+r$,~where 
$r=\max_{1 \leq i \leq m-1} v_{i}/v_{i+1}$. 
In this paper, we precisely analyze the behavior of {\sc greedy} and 
show~that~it is $(1+r)$-competitive 
for the case that $m$ queues do not necessarily 
have the same size. \medskip\\
{\bf Key Words:} Online Algorithms, Competitive Ratio, 
Buffer Management, Class Segregation, Quality of Service (QoS), 
Class of Service (CoS). 
%
\section{Introduction} \label{sec-introduction} 
%
Due to the burst growth of the Internet use, network traffic has 
increased year by year. This overloads networking systems and 
degrades the quality of communications, e.g., loss of bandwidth, 
packet~drops, delay of responses, etc. To overcome such degradation 
of the communication quality, the notion of Quality of Service (QoS) 
has received attention in practice, and is implemented by assigning 
nonnegative numerical values to packets to provide them with differentiated 
levels of service (priority).~Such~a packet value corresponds to the 
predefined Class of Service (CoS). In general, switches have several number of 
queues and each queue has a {\it buffer\/} to store arriving 
packets. Since network traffic changes frequently, switches need to 
control arriving packets to maximize the total values of transmitted packets, 
which is called {\it buffer management\/}. Basically, switches have no 
knowledge on the arrivals of packets in the future when it manages to 
control new packets arriving to the switches. So the decision made~by 
buffer management algorithm can be regarded as an {\it online algorithm\/}. 
In general,~the~performance~of online algorithms is measured by  
{\it competitive ratio\/} \cite{BE}. Online buffer management algorithms 
can be classified into two types of queue management  
(one is {\it preemptive\/} and the other is {\it nonpreemptive\/}).~Informally, 
we say that an online buffer management algorithm is 
preemptive if it is allowed~to~discard packets buffered in the queues on 
the arrival of new packets; nonpreemptive otherwise~(i.e.,~all packets 
buffered in the queues will be eventually transmitted). 
%
\subsection{Multi-Queue Buffer Management} \label{subsec-buffer}
%
In this paper, we focus on a multi-queue model  in which packets 
of different values are segregated~in different queues (see, e.g., \cite{CS}, 
\cite{LL}). Our model consists of $m$ packet values and $m$ 
queues\footnote[1]{~In general, we can consider a model of $m$ packet 
values and $n$ queues (with $m \neq n$), but in this paper, we deal with 
only a model of $m$ packet values and $m$ queues.}. 
Let~${\cal V}=\{v_{1},v_{2},\ldots,v_{m}\}$ be the set of $m$ nonnegative 
{\it packet values\/}, where $0<v_{1}<v_{2}<\cdots<v_{m}$, 
and~let~${\cal Q}=\{Q_{1},Q_{2},\ldots,Q_{m}\}$ be the set of $m$ 
queues. A packet of value $v_{j} \in  {\cal V}$ is referred to as a 
{\it $v_{j}$-packet\/},~and~a queue storing $v_{j}$-packets is referred to 
as a $v_{j}$-queue. 
Without loss of generality, we assume that~$Q_{j} \in {\cal Q}$ 
is a $v_{j}$-queue for each $j \in [1,m]$\footnote[2]{~For 
any pair of integers $a \leq b$, let $[a,b]= \{a,a+1,\ldots,b\}$. }. 
Each $Q_{j} \in {\cal Q}$ has a capacity $B_{j}\geq 1$,  i.e., each 
$Q_{j} \in {\cal Q}$ can store~up~to $B_{j} \geq 1$ packets. 
Since all packets buffered in each queue $Q_{j} \in {\cal Q}$ have the 
same value $v_{j} \in {\cal V}$,~the~order of transmitting packets buffered 
in queue $Q_{j} \in {\cal Q}$ is irrelevant. 

For convenience, we assume that time is discretized into slot of unit length. 
Packets arrive~over~time and each arriving packet is assigned with a 
(nonintegral) arrival time, a value $v_{j} \in {\cal V}$, and its destination queue 
$Q_{j} \in {\cal Q}$ (as we have assumed, $Q_{j} \in {\cal Q}$ is a $v_{j}$-queue). 
Let $\sigma$ be a sequence~of~{\it arrive\/}~events~and 
{\it send\/}  events, where an arrive event corresponds to the arrival 
of~a~new~packet~and~a~send~event~corresponds 
to the transmission of a packet 
buffered in queues at integral time (i.e., the end of~time~slot). 
An online (multi-queue) buffer 
management algorithm {\sc alg} consists of two~phases:~one~is an 
{\it admission\/} phase and the other is a {\it scheduling\/} phase.
In the admission phase, {\sc alg} must 
decide~on the arrival~of~a packet whether to accept or reject the packet 
with no knowledge on the future~arrivals~of~packets~(if~{\sc alg} is 
preemptive, then it may discard packets buffered in queues in the admission 
phase).~In~the scheduling 
phase, {\sc alg} chooses one of the nonempty 
queues at send event~and~exactly~one~packet~is~transmitted~out of the 
chosen queue. 
Since all packets buffered in the same queue have 
the same value, preemption~does not make sense in our model. Thus a packet 
accepted must eventually be transmitted. 

We say that an (online and offline) algorithm is {\it diligent\/} if 
(1) it must accept a packet arriving~to~its destination queue when the 
destination queue has vacancies, and 
(2) it must transmit a packet when~it has nonempty queues. 
It is not difficult to see that any nondiligent (online and offline) 
algorithm~can~be transformed to a 
diligent (online and offline) algorithm without decreasing its benefit 
(sum of~values~of transmitted packets). Thus in this paper, 
we focus on only diligent algorithms. 
%
\subsection{Main Results} \label{subsec-main}
%
Al-Bawani and Souza \cite[Theorem 2.2]{ABS2} presented an online 
multi-queue 
buffer management algorithm {\sc greedy} and showed that 
it is $(1+r)$-competitive 
for the case that $m$ queues have the same size, where 
\[
r = \max_{i \in [1,m-1]} \frac{v_{i}}{v_{i+1}}.
\]
In this paper, we remove the restriction that $m$ queue have the same size 
and show that the competitive ratio of 
{\sc greedy}~is~$1+r$ 
for the case that $m$ queues do not necessarily have 
the~same~size~(see~Theorem \ref{thm-main-1}). In addition, 
we construct a {\it bad\/} sequence $\sigma$ of events to 
show that the competitive ratio of {\sc greedy} is at least $1+r$ 
for the case that $m$ queues do not necessarily have the same size (see 
Theorem \ref{thm-bad}). 
%
\subsection{Related Works} \label{subsec-related}
%
The competitive analysis for the buffer management policies for switches 
were initiated by Aiello~et~al. \cite{AMRR}, Mansour et al. \cite{MPSL}, and 
Kesselman et al. \cite{Ketal}, and the extensive studies have been 
made~for~several models (for comprehensive surveys, see, e.g., 
\cite{A},\cite{ES},\cite{J},\cite{C},\cite{G}). 

The model we deal with in this paper can be regarded as the generalization of 
unit-valued model, where  the switches consist of $m$ queues 
of the same buffer size $B\geq 1$ and all packets have unit~value,~i.e., $v_{1}=v_{2}=\cdots=v_{m}$. The following tables summarize the known results (see 
Tables \ref{tab-det-cr} and \ref{tab-rand-cr}). 
\begin{table*}[htb]
\caption{Deterministic Competitive Ratio 
(Unit-Valued Multi-Queue Model)} \label{tab-det-cr}
\begin{center}
\def\arraystretch{1.2}
\begin{tabular}{|c|c||c|c|} \hline
\multicolumn{2}{|c||}{\makebox[5.0cm][c]{Upper Bound}} & 
\multicolumn{2}{c|}{\makebox[5.0cm][c]{Lower Bound}}\\  \hline
\begin{tabular}{cr}
\makebox[2.5cm][c]{2} & \makebox[1.0cm][r]{\cite{AR}}\\
\makebox[2.5cm][c]{1.889}  & \makebox[1.0cm][r]{\cite{AS}}\\
\makebox[2.5cm][c]{1.857} & \makebox[1.0cm][r]{\cite{AS}}\\
\makebox[2.5cm][c]{$\frac{e}{e-1}\approx 1.582$} & 
\makebox[1.0cm][r]{\cite{AL}}
\end{tabular} & 
\begin{tabular}{c}
---\\
\makebox[1.5cm][c]{$m \gg B$}\\
\makebox[1.5cm][c]{$B=2$}\\
\makebox[1.5cm][c]{large $B$}
\end{tabular} & 
\begin{tabular}{cr}
\makebox[2.5cm][c]{$2-1/m$} & \makebox[1.0cm][r]{\cite{AR}}\\
\makebox[2.5cm][c]{$1.366-\Theta(1/m)$} & \makebox[1.0cm][r]{\cite{AR}}\\
\makebox[2.5cm][c]{$\frac{e}{e-1}\approx 1.582$} & 
\makebox[1.0cm][r]{\cite{AS}}
\end{tabular} & 
\begin{tabular}{c}
\makebox[1.5cm][c]{$B=1$}\\
\makebox[1.5cm][c]{$B\geq 1$}\\
---
\end{tabular}\\ \hline
\end{tabular}
\end{center}
\caption{Randomized Competitive Ratio 
(Unit-Valued Multi-Queue Model)} \label{tab-rand-cr}
\begin{center}
\def\arraystretch{1.2}
\begin{tabular}{|c|c||c|c|} \hline
\multicolumn{2}{|c||}{\makebox[5.0cm][c]{Upper Bound}} & 
\multicolumn{2}{c|}{\makebox[5.0cm][c]{Lower Bound}}\\  \hline
\begin{tabular}{cr}
\makebox[2.5cm][c]{$\frac{e}{e-1}\approx 1.582$} & 
\makebox[1.0cm][r]{\cite{AR}}\\
\makebox[2.5cm][c]{1.231} & \makebox[1.0cm][r]{\cite{BM}}
\end{tabular} & 
\begin{tabular}{c}
\makebox[1.5cm][c]{$B > \log m$}\\
\makebox[1.5cm][c]{$m=2$}
\end{tabular} & 
\begin{tabular}{cr}
\makebox[2.5cm][c]{$1.46-\Theta(1/m)$}  & 
\makebox[1.0cm][r]{\cite{AR}}\\
\makebox[2.5cm][c]{1.4659}  & \makebox[1.0cm][r]{\cite{AS}}\\
\makebox[2.5cm][c]{1.231}  & \makebox[1.0cm][r]{\cite{AS}}\\
\makebox[2.5cm][c]{$\frac{e}{e-1}\approx 1.582$}  & 
\makebox[1.0cm][r]{\cite{B}}
\end{tabular} & 
\begin{tabular}{c}
\makebox[1.5cm][c]{$B=1$}\\
\makebox[1.5cm][c]{large $m$}\\
\makebox[1.5cm][c]{$m=2$}\\
\makebox[1.5cm][c]{---}\\
\end{tabular}\\ \hline
\end{tabular}
\end{center}
\end{table*}
On~the other hand, the model we deal with in this paper can be regarded as 
a special case of the general~$m$-valued multi-queue model, where each of $m$ 
queues can buffer at most $B$ packets~of~different values. 
For the preemptive multi-queue buffer management, 
Azar and Richter~\cite{AR}~showed~a~$(4+2\ln \alpha)$-competitive algorithm 
for the general $m$-valued case (packet values lie between 
1 and $\alpha$)~and~a 2.6-competitive~algo\-rithm for the two-valued case 
(packet values are $v_{1}<v_{2}$, where $v_{1}=1$ and $v_{2}=\alpha$). 
For the general~$m$-valued case, Azar and Righter \cite{AR2} proposed a more  
efficient algorithm {\sc transmit-largest head} ({\sc tlh}) that is 3-competitive, 
which is shown to be $(3-1/\alpha)$-competitive by Itoh and Takahashi \cite{IT}. 
%
\section{Preliminaries} \label{sec-preliminary}
%
\subsection{Notations and Terminologies} \label{subsec-notation}
%
Let $\sigma$ be a sequence of arrive and send 
events. Note that an arrive event corresponds~to~the~arrival~of a new 
packet (at nonintegral time) and a send event corresponds to the 
transmission~of~a~packet~buffered in queues at integral time. The 
algorithm {\sc greedy} works as follows: At send event, 
{\sc greedy}~transmits~a packet from the nonempty queue with the highest packet 
value\footnote[3]{~Since $Q_{j} \in {\cal Q}$ is a $v_{j}$-queue, such a 
nonempty queue with highest packet value is {\it unique\/} if it exists.}, 
i.e., {\sc greedy} transmits a $v_{h}$-packet if $v_{h}$-queue is nonempty 
and all $v_{\ell}$-queues are empty for $\ell \in [h+1,m]$. At arrive event, 
{\sc greedy} accepts packets in its destination queue until the corresponding 
queue becomes full. 

For an {\it online\/} algorithm {\sc alg} and a sequence $\sigma$, 
we use $\msc{alg}(\sigma)$ to denote the 
{\it benefit\/} of the algorithm {\sc alg} on the sequence $\sigma$, i.e., 
the sum of values of packets transmitted by {\sc alg}  
on $\sigma$. For a sequence~$\sigma$,~we 
also use $\msc{opt}(\sigma)$ to denote the {\it benefit\/} of the 
{\it optimal offline\/} algorithm {\sc opt} on the sequence 
$\sigma$,~i.e.,~the 
sum of values of packets transmitted by 
{\sc opt} that~knows~the~entire~sequence $\sigma$ in advance. 
For $c \geq 1$,~we say that an online algorithm~{\sc alg}~is~$c$-competitive 
if~$\msc{opt}(\sigma)/\msc{alg}(\sigma) \leq c$ 
for any~sequence $\sigma$. 
Thus our goal is to design an efficient (deterministic) online 
algorithm~{\sc alg}~that minimizes 
$\msc{opt}(\sigma)/\msc{alg}(\sigma)$ for any 
sequence $\sigma$. 
For~a~sequence~$\sigma$,~let~$A_{j}(\sigma)$~and 
$A_{j}^{*}(\sigma)$~be~the~total~number of $v_{j}$-packets 
accepted by {\sc greedy} and {\sc opt} until~the~end~of~the~sequence~$\sigma$,~respectively. 
When $\sigma$ is clear from the context, 
we simply denote $A_{j}$ and $A_{j}^{*}$ instead of $A_{j}(\sigma)$ and 
$A_{j}^{*}(\sigma)$, respectively. 
%
\subsection{Overview for \large{GREEDY}} \label{subsec-overview}
%
For the case that $B_{j}=B$ for each $j \in [1,m]$, 
Al-Bawani and Souza \cite{ABS2} derived the following lemmas~and 
showed that the competitive ratio of {\sc greedy} is $1+r$
\cite[Theorem 2.2]{ABS2}, where 
\[
r = \max_{i \in [1, m-1]} \frac{v_{i}}{v_{i+1}}. 
\]
\begin{lemma}[\mbox{\cite[Lemma 2.3]{ABS2}}] \label{lemma-2.3}
$A_{m}^{*}=A_{m}$.
\end{lemma}
\begin{lemma}[\mbox{\cite[Lemma 2.4]{ABS2}}] \label{lemma-2.4}
For any $i \in [1,m-1]$, $\sum_{j=i}^{m-1} (A_{j}^{*}-A_{j}) \leq 
\sum_{j=i+1}^{m} A_{j}$. 
\end{lemma}
\begin{lemma}[\mbox{\cite[Lemma 2.6]{ABS2}}] \label{lemma-2.6}
$\sum_{j=1}^{m-1} v_{j}(A_{j}^{*}-A_{j}) \leq \sum_{j=1}^{m-1} v_{j}A_{j+1}$. 
\end{lemma}
\begin{lemma}[\mbox{\cite[Lemma 2.7]{ABS2}}] \label{lemma-2.7}
$\sum_{j=1}^{m-1} v_{j}A_{j+1} \left / \sum_{j=1}^{m-1}v_{j+1}A_{j+1} 
\leq r \right.$. 
\end{lemma}
In fact, the competitive ratio of the algorithm {\sc greedy} can be derived 
as follows: 
\[
\frac{\msc{opt}(\sigma)}{\msc{greedy}(\sigma)} = 
\frac{\sum_{j=1}^{m} v_{j} A_{j}^{*}}{\sum_{j=1}^{m} v_{j}A_{j}}
= 1 + \frac{\sum_{j=1}^{m-1} v_{j} (A_{j}^{*}-A_{j})}{\sum_{j=1}^{m} v_{j}A_{j}}
\leq 1+\frac{\sum_{j=1}^{m-1} v_{j} A_{j+1}}{\sum_{j=1}^{m-1} v_{j+1}A_{j+1}}
\leq 1+r,
\]
where the second equality follows from Lemma \ref{lemma-2.3}, 
the first inequality follows from Lemma \ref{lemma-2.6},  and~the 
second inequality follows from Lemma \ref{lemma-2.7}. 

Lemmas \ref{lemma-2.3} and \ref{lemma-2.7} hold unless 
$B_{j}=B$ for each $j \in [1,m]$. On the other hand, 
Lemma~\ref{lemma-2.6}~immediately follows from Lemma \ref{lemma-2.4}, 
however, Lemma \ref{lemma-2.4} 
is shown only when $B_{j}=B$ for each~$j \in [1,m]$.~So 
for each $i \in [1,m-1]$, if 
$\sum_{j=i}^{m-1} (A_{j}^{*}-A_{j}) \leq \sum_{j=i+1}^{m} A_{j}$  holds 
for general $B_{j}$'s (i.e., it is not necessarily the case that 
$B_{j}=B$~for each $j \in [1,m]$), 
then we can show that the competitive ratio of {\sc greedy}~is 
$1+r$ for general $B_{j}$'s. In the following section, we extend 
Lemma \ref{lemma-2.4} to the case of general $B_{j}$'s,~which 
implies that the competitive ratio of the algorithm 
{\sc greedy} is $1+r$ for general $B_{j}$'s. 
%
\section{Upper Bounds} \label{sec-upper}
%
In this section, we show the following theorem. 
\begin{thm} \label{thm-main-1}
For $m$ packet values $0 < v_{1}<v_{2}<\cdots < v_{m}$, the competitive 
ratio of {\sc greedy}~is~$1+r$ for 
the case that $m$ queues do not necessarily have the same size, 
where $r =\max_{i \in [1,m-1]}v_{i}/v_{i+1}$. 
\end{thm}
As mentioned in Section \ref{subsec-overview}, the following lemma 
is essential to show Theorem \ref{thm-main-1} and is 
an extension of 
Lemma \ref{lemma-2.4} to the case that $m$ queues do not 
necessarily have the same size. 
\begin{lemma} \label{lemma-main}
For each $i \in [1,m-1]$, 
$\sum_{j=i}^{m-1} (A_{j}^{*}-A_{j}) \leq \sum_{j=i+1}^{m} A_{j}$  holds 
for general 
$B_{j}$'s $($i.e.,~it~is~not necessarily the case that 
$B_{j}=B$ for each $j \in [1,m])$. 
\end{lemma}
%
\subsection{Proof of Lemma \ref{lemma-main}} \label{subsec-proof-main}
%
For an arbitrarily fixed $i \in [1,m-1]$, let $V_{i} = \{v_{i},v_{i+1},\ldots,v_{m}\} \subseteq V$ 
and $\overline{V}_{i} = \{v_{1},v_{2},\ldots,v_{i-1}\} \subseteq V$. 
The notion of {\it time intervals\/} is defined as follows: 
A time interval $\msc{itv}$ ends with a send event~and~the next time interval 
starts with the first arrive event after the end of $\msc{itv}$.
We say that $\msc{itv}$~is~an~{\it $i$-red~interval\/} (or $r_{i}$-interval) if the 
value of any packet sent by {\sc greedy} during $\msc{itv}$ is in $V_{i}$, 
and~we~say~that~$\msc{itv}$ is an {\it $i$-green interval\/} 
(or $g_{i}$-interval) if the value of any packet 
sent by {\sc greedy} during~$\msc{itv}$~is~in~$\overline{V}_{i}$~or~$\msc{itv}$ contains send events at which {\sc greedy} sends no packets. 
Partition sequence $\sigma$ of events 
into $r_{i}$-intervals and $g_{i}$-intervals such that no two consecutive intervals 
are of the same color. 
It is easy to see that this partition is feasible. 
From the definition of {\sc greedy}, 
we have the following observation: 
\begin{observation}[\mbox{\cite[Observation 2.5]{ABS2}}] \label{observe-2.5}
For any $g_{i}$-interval and any $j \in [i,m]$, 
each $v_{j}$-queue~of~the~algorithm {\sc greedy} is 
empty and no $v_{j}$-packets arrive. 
\end{observation}
For any $j \in [i,m]$, 
let $A_{j}(\msc{itv})$ and $A_{j}^{*}(\msc{itv})$ be the total number 
of $v_{j}$-packets accepted by {\sc greedy}~and {\sc opt} in 
$\msc{itv}$, respectively. Let ${\cal R}_{i}$ be the set of all $r_{i}$-intervals. 
From Observation \ref{observe-2.5}, it follows that 
\[
A_{j}=\sum_{{\rm ITV} \in {\cal R}_{i}} A_{j}(\msc{itv});~~~~~
A_{j}^{*}=\sum_{{\rm ITV} \in {\cal R}_{i}} A_{j}^{*}(\msc{itv}).
\]
So it suffices to show Lemma \ref{lemma-main} for each 
$r_{i}$-interval 
$\msc{itv} \in {\cal R}_{i}$, i.e., 
for an arbitrarily fixed 
$\msc{itv} \in {\cal R}_{i}$, 
\begin{equation}
\sum_{j=i}^{m-1}\{A_{j}^{*}(\msc{itv})-A_{j}(\msc{itv})\}\leq \sum_{j=i+1}^{m} 
A_{j}(\msc{itv}). \label{eq-main-itv}
\end{equation}

Let $e_{1},e_{2},\ldots,e_{k}$ be 
events in  an arbitrarily fixed $\msc{itv} \in {\cal R}_{i}$. 
For {\sc greedy}, we use $\delta_{j}(e_{h})$ to denote the total number 
of $v_{j}$-packets sent by {\sc greedy} 
until the event $e_{h}$ of $\msc{itv}$ and 
$b_{j}(e_{h})$ to denote~the~number of packets contained 
in $v_{j}$-queue of {\sc greedy} 
just after the event $e_{h}$ of~$\msc{itv}$. 
For~{\sc opt},~we~use~$\delta_{j}^{+}(e_{h})$~to~denote the total 
number of $v_{j}$-packets sent by {\sc opt} until the event $e_{h}$ 
of $\msc{itv}$  and $b_{j}^{+}(e_{h})$~to~denote~the~number of packets 
contained  in $v_{j}$-queue of {\sc opt} just after the event 
$e_{h}$~of~$\msc{itv}$. 
Note that $\delta_{0}^{+}(e_{h})$ denotes~the total number of send events 
until the event $e_{h}$ at which {\sc opt} sends no packets. 
For each $j \in [i,m]$,~it~is immediate from 
Observation \ref{observe-2.5} that for {\sc greedy}, 
$\msc{itv}$ starts with $v_{j}$-queue 
empty and ends with~$v_{j}$-queue empty. 
Since no further $v_{j}$-packets arrive in 
$\msc{itv}$ after the (final) event $e_{k}$ of $\msc{itv}$, we have~that 
\begin{equation}
A_{j}(\msc{itv}) = \delta_{j}(e_{k})+b_{j}(e_{k})=\delta_{j}(e_{k}).  \label{eq-A_{j}-1}
\end{equation}

Let $r_{j}(e_{h}|\msc{g},\msc{o})$ 
be the total number of $v_{j}$-packets that are accepted 
by {\sc greedy} and {\sc opt} until the event $e_{h}$ of $\msc{itv}$, 
$r_{j}(e_{h}|\msc{g},\overline{\msc{o}})$ 
be the total number of $v_{j}$-packets that are 
accepted by {\sc greedy}~and~are~rejected by {\sc opt} until the event $e_{h}$ 
of $\msc{itv}$,  
$r_{j}(e_{h}|\overline{\msc{g}},\msc{o})$ 
be the total number of $v_{j}$-packets that~are~rejected by 
{\sc greedy} and are accepted by {\sc opt} until the event 
$e_{h}$ of $\msc{itv}$, 
and $r_{j}(e_{h}|\overline{\msc{g}},\overline{\msc{o}})$  
be the~total~number~of $v_{j}$-packets that are rejected 
by {\sc greedy} and {\sc opt} until the event $e_{h}$ of $\msc{itv}$. 
Then from the facts that 
$A_{j}(\msc{itv})=r_{j}(e_{k}|\msc{g},\msc{o})
+ r_{j}(e_{k}|\msc{g},\overline{\msc{o}})$ and 
$A_{j}^{*}(\msc{itv})=r_{j}(e_{k}|\msc{g},\msc{o})
+ r_{j}(e_{k}|\overline{\msc{g}},\msc{o})$, it follows that 
\begin{equation}
A_{j}(\msc{itv})-A_{j}^{*}(\msc{itv})  
= r_{j}(e_{k}|\msc{g},\overline{\msc{o}})-r_{j}(e_{k}|\overline{\msc{g}},\msc{o}). 
\label{eq-main'}
\end{equation}
Thus to prove that Equation (\ref{eq-main-itv}) holds, it suffices to show that 
\begin{eqnarray}
\varphi(e_{k}) & = & \sum_{j=i+1}^{m} A_{j}(\msc{itv}) + 
\sum_{j=i}^{m-1}\{A_{j}(\msc{itv})-A_{j}^{*}(\msc{itv})\}\nonumber\\
& = & \sum_{j=i+1}^{m} \delta_{j}(e_{k})+
\sum_{j=i}^{m-1} \{
r_{j}(e_{k}|\msc{g},\overline{\msc{o}})-r_{j}(e_{k}|\overline{\msc{g}},\msc{o})\} \geq 0, 
\label{eq-main-itv'}
\end{eqnarray}
where the second equality follows from Equations (\ref{eq-A_{j}-1}) and 
(\ref{eq-main'}). 

For each $j \in [1,m]$, we say that send event $e$ is {\it $(i,j)$-selecting\/} 
if {\sc greedy} sends a $v_{i}$-packet~and~{\sc opt} sends 
a $v_{j}$-packet  at the send event $e$, and say that send event $e$ is 
$(i,0)$-selecting if {\sc greedy} sends~a~$v_{i}$-packet and {\sc opt} 
sends not no packets at the send event $e$. 
For each $j \in [0,m]$, let $\Delta_{i,j}(e_{h})$ be 
the total number of $(i,j)$-selecting send events 
until the event $e_{h}$ of $\msc{itv}$. 
To show that Equation~(\ref{eq-main-itv'})~holds,~the~following 
claims are crucial. 
Let $N=\sum_{j=i}^{m} \delta_{j}(e_{k})$ be the total number of 
send events in $\msc{itv} \in {\cal R}_{i}$. 
\begin{claim} \label{claim-1}
For each $j \in [i,m-1]$, $r_{j}(e_{k}|\msc{g},\overline{\msc{o}})-
r_{j}(e_{k}|\overline{\msc{g}},\msc{o})\geq 
\Delta_{i,j}(e_{k})-\delta_{j}^{+}(e_{k})$.
\end{claim}
\begin{claim} \label{claim-2}
%
$N \geq \sum_{j=0}^{i-1} \Delta_{i,j}(e_{k}) + 
\sum_{j=i}^{m-1} \delta_{j}^{+}(e_{k})+\Delta_{i,m}(e_{k})$.
\end{claim}
The proofs of Claims \ref{claim-1} and \ref{claim-2}  are given in 
Sections \ref{subsubsec-proof-lemma-1} and 
\ref{subsubsec-proof-lemma-2}, respectively. 
From Claims~\ref{claim-1}~and \ref{claim-2}, 
we can immediately derive Equation (\ref{eq-main-itv'}) 
as follows: 
\begin{eqnarray}
\varphi(e_{k}) & = & 
\sum_{j=i+1}^{m} \delta_{j}(e_{k})+
\sum_{j=i}^{m-1} \{
r_{j}(e_{k}|\msc{g},\overline{\msc{o}})-
r_{j}(e_{k}|\overline{\msc{g}},\msc{o})\}\nonumber\\
& \geq & 
\sum_{j=i+1}^{m} \delta_{j}(e_{k})+
\sum_{j=i}^{m-1} \{
\Delta_{i,j}(e_{k})-\delta_{j}^{+}(e_{k})
\}\nonumber\\
& = & N - \delta_{i}(e_{k}) + \sum_{j=i}^{m-1} \{
\Delta_{i,j}(e_{k})-\delta_{j}^{+}(e_{k})
\}, \label{eq-1}
\end{eqnarray}
where the first inequality follows from Claim \ref{claim-1} and the second 
equality follows from the fact~that~$N=\sum_{j=i}^{m} \delta_{j}(e_{k})$. 
Note that $\delta_{i}(e_{k})=
\sum_{j=0}^{m} \Delta_{i,j}(e_{k})$. 
%
%
%
%
Then from Equation (\ref{eq-1}), it follows that 
\begin{eqnarray*}
\varphi(e_{k}) & \geq & N - \sum_{j=0}^{m} \Delta_{i,j}(e_{k})
+ \sum_{j=i}^{m-1} \{
\Delta_{i,j}(e_{k})-\delta_{j}^{+}(e_{k})\}\\
& = & N - 
\sum_{j=0}^{i-1} \Delta_{i,j}(e_{k})-
\Delta_{i,m}(e_{k}) 
- \sum_{j=i}^{m-1} \delta_{j}^{+}(e_{k})
\geq 0, 
\end{eqnarray*}
where the last inequality follows from Claim \ref{claim-2}. 
Thus this completes the proof of Lemma \ref{lemma-main}. 
%
\subsection{Proofs of Claims}  \label{subsec-proof}
%
\subsubsection{Proof of Claim \ref{claim-1}} 
\label{subsubsec-proof-lemma-1}
%
For  each $h \in [1,k]$, we use $\alpha_{j}(e_{h}) \geq 0$ to denote 
the {\it margin\/} of $v_{j}$-queue at the event $e_{h}$, i.e., 
\begin{equation}
\alpha_{j}(e_{h}) = \max \left\{0,b_{j}(e_{h}) - b_{j}^{+}(e_{h})\right\}=
\left\{
\begin{array}{ccl}
b_{j}(e_{h}) - b_{j}^{+}(e_{h}) & & b_{j}(e_{h}) > b_{j}^{+}(e_{h});\\
0 & & b_{j}(e_{h}) \leq b_{j}^{+}(e_{h}). 
\end{array}  \right. \label{eq-def-alpha}
\end{equation}
Note that $\alpha_{j}(e_{h}) \geq 0$ by definition. 
Since $b_{j}(e_{k})=0$ by Observation \ref{observe-2.5}, 
we have that $\alpha_{j}(e_{k})=0$.~Then to prove that 
$r_{j}(e_{k}|\msc{g},\overline{\msc{o}})-
r_{j}(e_{k}|\overline{\msc{g}},\msc{o})\geq 
\Delta_{i,j}(e_{k})-\delta_{j}^{+}(e_{k})$,~it suffices to show that for each~$h \in [1,k]$, 
\begin{equation}
r_{j}(e_{h}|\msc{g},\overline{\msc{o}})-
r_{j}(e_{h}|\overline{\msc{g}},\msc{o})\geq 
\Delta_{i,j}(e_{h})-\delta_{j}^{+}(e_{h})
+\alpha_{j}(e_{h}). 
\label{eq-3}
\end{equation}
For an arbitrarily fixed $j \in [i,m-1]$, 
we derive Equation (\ref{eq-3}) by induction on $h \in [1,k]$. 

{\sf Base Step:} From the definition of 
$\msc{itv} \in {\cal R}_{i}$, 
it follows that $e_{1}$ is arrive event, and from Observation 
\ref{observe-2.5}, it follows that 
$v_{\ell}$-queue of {\sc greedy} is empty just before the event $e_{1}$ 
for each $\ell \in [i,m]$. 
Assume~that a $v_{s}$-packet arrives at the event $e_{1}$. 
Let us consider the following cases: (a) $s=j$ and (b) $s \neq j$.

(a) $s=j$: 
Since $v_{j}$-queue of {\sc greedy} is empty just before the event 
$e_{1}$, {\sc greedy} accepts~a~$v_{j}$-packet at the event $e_{1}$. 
So it is obvious that $r_{j}(e_{1}|\msc{g},\overline{\msc{o}})\geq0$,  
$r_{j}(e_{1}|\overline{\msc{g}},\msc{o})=0$, 
and $b_{j}(e_{1})=1$.  
Since $e_{1}$ is arrive event, 
we have that  $\Delta_{i,j}(e_{1})=\delta_{j}^{+}(e_{1})=0$. 
We claim that $\alpha_{j}(e_{1})=0$. 
If {\sc opt} accepts a $v_{j}$-packet~at~the event $e_{1}$, 
then we have that $b_{j}^{+}(e_{1}) \geq 1 = b_{j}(e_{1})$, 
and if {\sc opt} rejects 
a $v_{j}$-packet at the event $e_{1}$, then~we have that 
$b_{j}^{+}(e_{1})=B_{j} \geq 1=b_{j}(e_{1})$. 
Thus in  Case (a), it follows that 
Equation (\ref{eq-3}) holds for $h=1$. 

(b) $s\neq j$:  Since $v_{j}$-queue of {\sc greedy} is empty just 
before the event $e_{1}$ and no $v_{j}$-packets arrive~at the 
event $e_{1}$, we have that 
$r_{j}(e_{1}|\msc{g},\overline{\msc{o}})=  
r_{j}(e_{1}|\overline{\msc{g}},\msc{o})=b_{j}(e_{1})=0$. 
From the fact that~$e_{1}$~is~arrive~event, it follows that 
$\Delta_{i,j}(e_{1})=\delta_{j}^{+}(e_{1})=0$. 
Since $b_{j}(e_{1})=0$,~we~have that $b_{j}^{+}(e_{1})\geq b_{j}(e_{1})$, i.e., 
$\alpha_{j}(e_{1})=0$. 
Thus in Case (b), it follows that 
Equation (\ref{eq-3}) holds for $h=1$. 

{\sf Induction Step:} For any $\ell \in [2,k]$, we assume that 
Equation (\ref{eq-3}) holds for $h=\ell-1$, i.e.,
\begin{equation}
r_{j}(e_{\ell-1}|\msc{g},\overline{\msc{o}})-
r_{j}(e_{\ell-1}|\overline{\msc{g}},\msc{o})\geq 
\Delta_{i,j}(e_{\ell-1})-\delta_{j}^{+}(e_{\ell-1})
+\alpha_{j}(e_{\ell-1}). \label{eq-4}
\end{equation}
For the event $e_{\ell}$, 
let us consider the 
following cases: (c) $e_{\ell}$ is arrive event and (d) 
$e_{\ell}$ is send event. 

(c) $e_{\ell}$ is arrive event: 
Assume that a $v_{s}$-packet arrives at the event $e_{\ell}$. 
Since $e_{\ell}$ is arrive event,~it~is immediate that 
$\Delta_{i,j}(e_{\ell})=\Delta_{i,j}(e_{\ell-1})$ and 
$\delta_{j}^{+}(e_{\ell})=\delta_{j}^{+}(e_{\ell-1})$.
If $s \neq j$,  then 
$r_{j}(e_{\ell}|\msc{g},\overline{\msc{o}})=
r_{j}(e_{\ell-1}|\msc{g},\overline{\msc{o}})$, 
$r_{j}(e_{\ell}|\overline{\msc{g}},\msc{o})=
r_{j}(e_{\ell-1}|\overline{\msc{g}},\msc{o})$, 
and $\alpha_{j}(e_{\ell})=\alpha_{j}(e_{\ell-1})$ hold. 
Thus from Equation (\ref{eq-4}), it follows~that~Equation (\ref{eq-3}) holds for $h=\ell$. So we assume that $s=j$ and 
let us consider the following cases: 
(c-1)~both {\sc greedy}  and {\sc opt} accept~the~$v_{j}$-packet; 
(c-2) both {\sc greedy}  and {\sc opt} reject the $v_{j}$-packet;  
(c-3) {\sc greedy} rejects 
and {\sc opt} accepts~the~$v_{j}$-packet; 
(c-4) {\sc greedy}~accepts 
and {\sc opt} rejects the $v_{j}$-packet. 

For Case (c-1), {\sc greedy} and {\sc opt} accept the $v_{j}$-packet 
at the event $e_{\ell}$. So 
we have that 
$r_{j}(e_{\ell}|\msc{g},\overline{\msc{o}})=
r_{j}(e_{\ell-1}|\msc{g},\overline{\msc{o}})$, 
$r_{j}(e_{\ell}|\overline{\msc{g}},\msc{o})=
r_{j}(e_{\ell-1}|\overline{\msc{g}},\msc{o})$, 
$b_{j}(e_{\ell})=b_{j}(e_{\ell-1})+1$,  and 
$b_{j}^{+}(e_{\ell})=b_{j}^{+}(e_{\ell-1})+1$. This implies that 
$\alpha_{j}(e_{\ell})=\alpha_{j}(e_{\ell-1})$. Thus from Equation (\ref{eq-4}), 
it follows that 
\begin{eqnarray*}
r_{j}(e_{\ell}|\msc{g},\overline{\msc{o}})-
r_{j}(e_{\ell}|\overline{\msc{g}},\msc{o}) & =& 
r_{j}(e_{\ell-1}|\msc{g},\overline{\msc{o}})-
r_{j}(e_{\ell-1}|\overline{\msc{g}},\msc{o})\\
& \geq & 
\Delta_{i,j}(e_{\ell-1})-\delta_{j}^{+}(e_{\ell-1})
+\alpha_{j}(e_{\ell-1})\\
& = & 
\Delta_{i,j}(e_{\ell})-\delta_{j}^{+}(e_{\ell})
+\alpha_{j}(e_{\ell}).
\end{eqnarray*}
For Case (c-2), 
{\sc greedy} and {\sc opt} 
reject the $v_{j}$-packet at the event $e_{\ell}$. Then we have that 
$r_{j}(e_{\ell}|\msc{g},\overline{\msc{o}})=
r_{j}(e_{\ell-1}|\msc{g},\overline{\msc{o}})$, 
$r_{j}(e_{\ell}|\overline{\msc{g}},\msc{o})=
r_{j}(e_{\ell-1}|\overline{\msc{g}},\msc{o})$, 
$b_{j}(e_{\ell})=b_{j}(e_{\ell-1})$,  and 
$b_{j}^{+}(e_{\ell})=b_{j}^{+}(e_{\ell-1})$. This immediately 
implies~that $\alpha_{j}(e_{\ell})=\alpha_{j}(e_{\ell-1})$. 
Thus from Equation (\ref{eq-4}), it follows that 
\begin{eqnarray*}
r_{j}(e_{\ell}|\msc{g},\overline{\msc{o}})-
r_{j}(e_{\ell}|\overline{\msc{g}},\msc{o}) & =& 
r_{j}(e_{\ell-1}|\msc{g},\overline{\msc{o}})-
r_{j}(e_{\ell-1}|\overline{\msc{g}},\msc{o})\\
& \geq & 
\Delta_{i,j}(e_{\ell-1})-\delta_{j}^{+}(e_{\ell-1})
+\alpha_{j}(e_{\ell-1})\\
& = & 
\Delta_{i,j}(e_{\ell})-\delta_{j}^{+}(e_{\ell})
+\alpha_{j}(e_{\ell}).
\end{eqnarray*}
For Case (c-3),  {\sc greedy} rejects and 
{\sc opt} accepts the $v_{j}$-packet at the event $e_{\ell}$. 
So it is easy~to~see~that $r_{j}(e_{\ell}|\msc{g},\overline{\msc{o}})=
r_{j}(e_{\ell-1}|\msc{g},\overline{\msc{o}})$, 
$r_{j}(e_{\ell}|\overline{\msc{g}},\msc{o})=
r_{j}(e_{\ell-1}|\overline{\msc{g}},\msc{o})+1$,  
$b_{j}(e_{\ell}) =b_{j}(e_{\ell-1})=B_{j}$, 
$b_{j}^{+}(e_{\ell}) =b_{j}^{+}(e_{\ell-1})+1 \leq B_{j}$, 
and 
$\alpha_{j}(e_{\ell-1})=b_{j}(e_{\ell-1})-b_{j}^{+}(e_{\ell-1})\geq 1$. 
This implies that 
\[
\alpha_{j}(e_{\ell})=b_{j}(e_{\ell})-b_{j}^{+}(e_{\ell}) = 
b_{j}(e_{\ell-1})-b_{j}^{+}(e_{\ell-1})-1 = \alpha_{j}(e_{\ell-1})-1 \geq 0. 
\]
Thus from Equation (\ref{eq-4}), it follows that 
\begin{eqnarray*}
r_{j}(e_{\ell}|\msc{g},\overline{\msc{o}})-
r_{j}(e_{\ell}|\overline{\msc{g}},\msc{o}) & = & 
r_{j}(e_{\ell-1}|\msc{g},\overline{\msc{o}})-
r_{j}(e_{\ell-1}|\overline{\msc{g}},\msc{o})-1\\
& \geq & 
\Delta_{i,j}(e_{\ell-1})-\delta_{j}^{+}(e_{\ell-1})
+\alpha_{j}(e_{\ell-1})-1\\
& = & 
\Delta_{i,j}(e_{\ell})-\delta_{j}^{+}(e_{\ell})
+\alpha_{j}(e_{\ell}).
\end{eqnarray*}
For Case (c-4), {\sc greedy} accepts 
and {\sc opt} rejects the $v_{j}$-packet at the event $e_{\ell}$. So  
it~is~immediate~to~see that $r_{j}(e_{\ell}|\msc{g},\overline{\msc{o}})=
r_{j}(e_{\ell-1}|\msc{g},\overline{\msc{o}})+1$, 
$r_{j}(e_{\ell}|\overline{\msc{g}},\msc{o})=
r_{j}(e_{\ell-1}|\overline{\msc{g}},\msc{o})$,  
$b_{j}(e_{\ell}) =b_{j}(e_{\ell-1})+1 \leq B_{j}$, 
$b_{j}^{+}(e_{\ell}) =b_{j}^{+}(e_{\ell-1})=B_{j}$, 
and 
$b_{j}(e_{\ell-1})-b_{j}^{+}(e_{\ell-1})\leq -1$. 
This implies that 
\[
b_{j}(e_{\ell})-b_{j}^{+}(e_{\ell}) = 
b_{j}(e_{\ell-1})+1-b_{j}^{+}(e_{\ell-1}) \leq 0, 
\] 
and we have that 
$\alpha_{j}(e_{\ell-1})=\alpha_{j}(e_{\ell})=0$. 
Thus from Equation (\ref{eq-4}), it follows that 
\begin{eqnarray*}
r_{j}(e_{\ell}|\msc{g},\overline{\msc{o}})-
r_{j}(e_{\ell}|\overline{\msc{g}},\msc{o}) & = & 
r_{j}(e_{\ell-1}|\msc{g},\overline{\msc{o}})+1-
r_{j}(e_{\ell-1}|\overline{\msc{g}},\msc{o})\\
& \geq & 
\Delta_{i,j}(e_{\ell-1})-\delta_{j}^{+}(e_{\ell-1})
+\alpha_{j}(e_{\ell-1})+1\\
& > & 
\Delta_{i,j}(e_{\ell})-\delta_{j}^{+}(e_{\ell})
+\alpha_{j}(e_{\ell}).
\end{eqnarray*}
Hence in Case (c), we have that Equation ({\ref{eq-3}) holds for $h = \ell$. 

(d) $e_{\ell}$ is send event: Let 
$v_{x}$ and $v_{y}$ be the values of packets sent by {\sc greedy} 
and {\sc opt} at the event~$e_{\ell}$, respectively. We consider the following 
cases: 
(d-1) $y\neq j$; (d-2) $y=j$ and $x\neq i$; (d-3)~$y=j$~and~$x=i$. 
Since $e_{\ell}$ is send~event, we have that 
$r_{j}(e_{\ell}|\msc{g},\overline{\msc{o}})=
r_{j}(e_{\ell-1}|\msc{g},\overline{\msc{o}})$ and 
$r_{j}(e_{\ell}|\overline{\msc{g}},\msc{o}) = 
r_{j}(e_{\ell-1}|\overline{\msc{g}},\msc{o})$. 

For Case (d-1), {\sc opt} does not send a $v_{j}$-packet 
at the event $e_{\ell}$. It is obvious that 
$b_{j}^{+}(e_{\ell})=b_{j}^{+}(e_{\ell-1})$, 
$\Delta_{i,j}(e_{\ell})=\Delta_{i,j}(e_{\ell-1})$, 
$\delta_{j}^{+}(e_{\ell})=\delta_{j}^{+}(e_{\ell-1})$,  and 
$b_{j}(e_{\ell}) \leq b_{j}(e_{\ell-1})$.~This implies that 
$\alpha_{j}(e_{\ell}) \leq \alpha_{j}(e_{\ell-1})$. 
Thus from Equation (\ref{eq-4}), it follows that 
\begin{eqnarray*}
r_{j}(e_{\ell}|\msc{g},\overline{\msc{o}})-
r_{j}(e_{\ell}|\overline{\msc{g}},\msc{o}) & = & 
r_{j}(e_{\ell-1}|\msc{g},\overline{\msc{o}})-
r_{j}(e_{\ell-1}|\overline{\msc{g}},\msc{o})\\
& \geq & 
\Delta_{i,j}(e_{\ell-1})-\delta_{j}^{+}(e_{\ell-1})
+\alpha_{j}(e_{\ell-1})\\
& \geq & 
\Delta_{i,j}(e_{\ell})-\delta_{j}^{+}(e_{\ell})
+\alpha_{j}(e_{\ell}).
\end{eqnarray*}
For Case (d-2), {\sc opt} sends a $v_{j}$-packet at the event $e_{\ell}$. 
It is obvious that $\delta_{j}^{+}(e_{\ell})
=\delta_{j}^{+}(e_{\ell-1})+1$,~$b_{j}^{+}(e_{\ell})=b_{j}^{+}(e_{\ell-1})-1$, and $b_{j}(e_{\ell}) \leq b_{j}(e_{\ell-1})$, and 
it follows that $\alpha_{j}(e_{\ell})\leq \alpha_{j}(e_{\ell-1})+1$. 
Since {\sc greedy}~does~not send a $v_{i}$-packet 
at the event $e_{\ell}$, we have that 
$\Delta_{i,j}(e_{\ell})=\Delta_{i,j}(e_{\ell-1})$. 
From Equation (\ref{eq-4}), it follows~that 
\begin{eqnarray*}
r_{j}(e_{\ell}|\msc{g},\overline{\msc{o}})-
r_{j}(e_{\ell}|\overline{\msc{g}},\msc{o}) & = & 
r_{j}(e_{\ell-1}|\msc{g},\overline{\msc{o}})-
r_{j}(e_{\ell-1}|\overline{\msc{g}},\msc{o})\\
& \geq & 
\Delta_{i,j}(e_{\ell-1})-\delta_{j}^{+}(e_{\ell-1})
+\alpha_{j}(e_{\ell-1})\\
& = & 
\Delta_{i,j}(e_{\ell})-\delta_{j}^{+}(e_{\ell})+1
+\alpha_{j}(e_{\ell-1})\\
& \geq & 
\Delta_{i,j}(e_{\ell})-\delta_{j}^{+}(e_{\ell})
+\alpha_{j}(e_{\ell}). 
\end{eqnarray*}
For Case (d-3), we further consider the following cases: 
(d-3.1) $i=j$ and (d-3.2)~$i<j$.~For~Case~(d-3.1), 
both {\sc greedy} and {\sc opt} 
sends a $v_{j}$-packet at the event $e_{\ell}$. Then it is immediate 
that $b_{j}(e_{\ell})=b_{j}(e_{\ell-1})-1$, 
$b_{j}^{+}(e_{\ell})=b_{j}^{+}(e_{\ell-1})-1$, 
$\delta_{j}^{+}(e_{\ell})=\delta_{j}^{+}(e_{\ell-1})+1$, 
and $\Delta_{i,j}(e_{\ell})=\Delta_{i,j}(e_{\ell-1})+1$. 
This~implies~that~$\alpha_{j}(e_{\ell})=\alpha_{j}(e_{\ell-1})$ by definition. 
Thus from Equation (\ref{eq-4}), it follows that 
\begin{eqnarray*}
r_{j}(e_{\ell}|\msc{g},\overline{\msc{o}})-
r_{j}(e_{\ell}|\overline{\msc{g}},\msc{o}) & = & 
r_{j}(e_{\ell-1}|\msc{g},\overline{\msc{o}})-
r_{j}(e_{\ell-1}|\overline{\msc{g}},\msc{o})\\
& \geq & 
\Delta_{i,j}(e_{\ell-1})-\delta_{j}^{+}(e_{\ell-1})
+\alpha_{j}(e_{\ell-1})\\
& = & 
\Delta_{i,j}(e_{\ell})-1-\delta_{j}^{+}(e_{\ell-1})
+\alpha_{j}(e_{\ell})\\
& = & 
\Delta_{i,j}(e_{\ell})-\delta_{j}^{+}(e_{\ell})
+\alpha_{j}(e_{\ell}). 
\end{eqnarray*}
For Case (d-3.2), {\sc greedy} sends a $v_{i}$-packet and 
{\sc opt} sends a $v_{j}$-packet at the event $e_{\ell}$. 
It is immediate that $\delta_{j}^{+}(e_{\ell})=
\delta_{j}^{+}(e_{\ell-1})+1$ 
and $\Delta_{i,j}(e_{\ell})=\Delta_{i,j}(e_{\ell-1})+1$. 
Since $i < j$, we have that 
$b_{j}(e_{\ell-1})=0$~by~defi\-nition 
(if $b_{j}(e_{\ell-1})>0$, 
then $v_{i}$ is not the highest packet value among the packets residing 
in~queues~just after the event $e_{\ell-1}$ and 
{\sc greedy} does not send a $v_{i}$-packet at the event $e_{\ell}$). 
So~it~follows~that~$b_{j}(e_{\ell})=b_{j}(e_{\ell-1})=0$ and this implies that 
$\alpha_{j}(e_{\ell})=0 \leq \alpha_{j}(e_{\ell-1})$. 
Thus from Equation (\ref{eq-4}), it follows that 
\begin{eqnarray*}
r_{j}(e_{\ell}|\msc{g},\overline{\msc{o}})-
r_{j}(e_{\ell}|\overline{\msc{g}},\msc{o}) & = & 
r_{j}(e_{\ell-1}|\msc{g},\overline{\msc{o}})-
r_{j}(e_{\ell-1}|\overline{\msc{g}},\msc{o})\\
& \geq & 
\Delta_{i,j}(e_{\ell-1})-\delta_{j}^{+}(e_{\ell-1})
+\alpha_{j}(e_{\ell-1})\\
& \geq & 
\Delta_{i,j}(e_{\ell})-1-\delta_{j}^{+}(e_{\ell-1})
+\alpha_{j}(e_{\ell})\\
& = & 
\Delta_{i,j}(e_{\ell})-\delta_{j}^{+}(e_{\ell})
+\alpha_{j}(e_{\ell}). 
\end{eqnarray*}
Hence in Case (d), we have that Equation ({\ref{eq-3}) holds for $h = \ell$. 
%
\subsubsection{Proof of Claim \ref{claim-2}} 
\label{subsubsec-proof-lemma-2}
%
Since $\Delta_{i,j}(e_{k})$ is the total number~of~$(i,j)$-selecting send 
events in $\msc{itv}$ and $\delta_{j}^{+}(e_{k})$ is the 
total number~of $v_{j}$-packets sent by {\sc opt} in $\msc{itv}$, 
we have that $\Delta_{i,j}(e_{k}) \leq 
\delta_{j}^{+}(e_{k})$ for each $j \in [0,m]$. 
Thus it follows~that 
\begin{eqnarray*}
\sum_{j=0}^{i-1} \Delta_{i,j}(e_{k}) 
+ \sum_{j=i}^{m-1} \delta_{j}^{+}(e_{k})+\Delta_{i,m}(e_{k}) 
& \leq & 
\sum_{j=0}^{i-1} \delta_{j}^{+}(e_{k})
+ \sum_{j=i}^{m-1} \delta_{j}^{+}(e_{k})+
\delta_{m}^{+}(e_{k})\\
& = & \sum_{j=0}^{m} \delta_{j}^{+}(e_{k}) =  N, 
\end{eqnarray*}
where the second equality follows from the fact that 
$N$ is the total number of send events in $\msc{itv}$. 
%
\section{Lower Bounds} \label{sec-lower}
%
In this section, we derive lower bounds for the competitive ratio of the algorithm 
{\sc greedy},~which~shows that the competitive ratio of {\sc greedy} 
cannot improve any more. 
\begin{thm} \label{thm-bad}
For $m$ packet values $0 < v_{1}<v_{2}<\cdots < v_{m}$ and 
any $\varepsilon > 0$, 
the competitive~ratio~of the algorithm {\sc greedy} cannot be less than 
$1+r - \varepsilon$ 
for the case that $m$ queues do not necessarily~have the same size,  
where $r =\max_{ i \in [1,m-1]} v_{i}/v_{i+1}$. 
\end{thm}
{\bf Proof:} To derive lower bounds for the competitive ratio of 
{\sc greedy} for the case that $m$ queues do not necessarily 
have the same size, 
define a sequence $\sigma$ as follows:~The sequence 
$\sigma$~consists~of~$m$ phases.~The 
phase $P_{1}$ includes $B_{m}$ time slots. 
In the 1st time slot of the phase~$P_{1}$, $B_{1}$ copies of 
$v_{1}$-packet arrive,~$B_{2}$ copies of $v_{2}$-packet arrive, 
$\ldots$, and $B_{m}$ copies of $v_{m}$-packet arrive. 
For~each~$i \in [2,B_{m}]$,~a~$v_{m-1}$-packet arrives 
in the $i$th time slot of the phase $P_{1}$. 
For each $j \in [2,m]$, the phase $P_{j}$ includes $B_{m+1-j}$~time~slots. 
In the 1th time slot of the phase $P_{j}$, a $v_{m+1-j}$-packet arrives. 
For each $i \in [2,B_{m+1-j}]$, a $v_{m-j}$-packet arrives 
in the $i$th time slot of the phase $P_{j}$. 
Regard $v_{0}$-packet as a null packet and 
this implies~that~no packets arrive 
in the $i$th time slot of 
the phase $P_{m}$  with $i \in [2,B_{1}]$. 

On the sequence $\sigma$, the behavior 
of {\sc greedy} is given in Figure \ref{fig-greedy}. 
From the definition of {\sc greedy},~it~is immediate that 
$B_{m}$ copies of $v_{m}$-packets are sent in the phase $P_{1}$, 
$B_{m-1}$ copies of $v_{m-1}$-packets~are~sent in the phase $P_{2}$, 
$\ldots$, and $B_{1}$ copies of $v_{1}$-packets are sent in the phase $P_{m}$. 
For the queues of {\sc greedy},~we observe that 
for each $j \in [1,m]$, 
$v_{1}$-queue, $\ldots$, $v_{m-j}$-queue are 
full and $v_{m-j+1}$-queue, $\ldots$, $v_{m}$-queue~are empty 
at the end of the phase $P_{j}$. 
Thus for the benefit $\msc{greedy}(\sigma)$, it follows that 
\[
\msc{greedy}(\sigma) =  B_{1}v_{1}+B_{2}v_{2}+\cdots + 
B_{m-1}v_{m-1}+B_{m}v_{m}.  
\]
We consider the following offline algorithm {\sc adv} 
(on the sequence $\sigma$, the behavior of {\sc adv} 
is given~in~Figure \ref{fig-adv}). 
For each $j \in [1,m-1]$ and each $i \in [1,B_{m+1-j}]$,~{\sc adv} 
sends a $v_{m-j}$-packet at the end 
of~the~$i$th time slot of the phase $P_{j}$. 
For the queues of {\sc adv}, we observe that 
for each $j \in [1,m]$, every queue~is~full just before the send event 
in the 1st time slot of the phase $P_{j}$. 
Then it follows~that~{\sc adv}~sends~$B_{m}$~copies of $v_{m-1}$-packets 
in the phase $P_{1}$, 
$B_{m-1}$ copies of $v_{m-2}$-packets in the phase $P_{2}$, 
$\ldots$, and~$B_{2}$~copies~of~$v_{1}$-packets in the phase $P_{m-1}$. 
In particular, we have that 
just after the arrive event $e_{*}$ in the~1st~time~slot of the 
phase~$P_{m}$, 
every queue of {\sc adv} is full 
and no further packets arrive. 
This implies that after~the~arrive event $e_{*}$ 
in the 1st time slot of the phase~$P_{m}$,~{\sc adv}~sends~$B_{1}$~copies 
of $v_{1}$-packets,~$B_{2}$~copies~of~$v_{2}$-packets, 
$\ldots$, and $B_{m}$ copies of $v_{m}$-packets. 
Thus for the benefit $\msc{opt}(\sigma)$, 
we have that 
\begin{eqnarray}
\msc{opt}(\sigma) & \geq & \msc{adv}(\sigma) 
= (B_{1}+B_{2})v_{1}+(B_{2}+B_{3})v_{2}+\cdots+
(B_{m-1}+B_{m})v_{m-1}+B_{m}v_{m}\nonumber\\
& = & B_{1}v_{1}+B_{2}(v_{1}+v_{2})+
B_{3}(v_{2}+v_{3})+\cdots+
B_{m}(v_{m-1}+v_{m}). \label{eq-o}
\end{eqnarray}

Assume that $r=v_{\ell}/v_{\ell+1}=\max_{i \in [1,m-1]} v_{i}/v_{i+1}$
for some $\ell \in [1,m-1]$. 
Note that 
\begin{eqnarray*}
\frac{\msc{opt}(\sigma)}{\msc{greedy}(\sigma)} & \geq & 
\frac{\msc{adv}(\sigma)}{\msc{greedy}(\sigma)}\\
& = & 
\frac{
\frac{B_{1}}{B_{\ell+1}}v_{1}+
\frac{B_{2}}{B_{\ell+1}}(v_{1}+v_{2})+
\cdots +
(v_{\ell}+v_{\ell+1})
+\cdots+
\frac{B_{m}}{B_{\ell+1}}(v_{m-1}+v_{m})}{
\frac{B_{1}}{B_{\ell+1}}v_{1}+
\frac{B_{2}}{B_{\ell+1}}v_{2}+
\cdots+v_{\ell+1}+\cdots+
\frac{B_{m}}{B_{\ell+1}}v_{m}}. 
%
\end{eqnarray*}
For each $j \in [1,m]\setminus \{\ell+1\}$,~set $B_{j}=1$. Then we have that 
\begin{eqnarray*}
\lefteqn{\lim_{B_{\ell+1}\to \infty} \frac{
\frac{B_{1}}{B_{\ell+1}}v_{1}+
\frac{B_{2}}{B_{\ell+1}}(v_{1}+v_{2})+
\cdots +
(v_{\ell}+v_{\ell+1})
+\cdots+
\frac{B_{m}}{B_{\ell+1}}(v_{m-1}+v_{m})}{
\frac{B_{1}}{B_{\ell+1}}v_{1}+
\frac{B_{2}}{B_{\ell+1}}v_{2}+
\cdots+v_{\ell+1}+\cdots+
\frac{B_{m}}{B_{\ell+1}}v_{m}}}\\
& = & 
\lim_{B_{\ell+1}\to \infty} \frac{
\frac{1}{B_{\ell+1}}v_{1}+
\frac{1}{B_{\ell+1}}(v_{1}+v_{2})+
\cdots +
(v_{\ell}+v_{\ell+1})
+\cdots+
\frac{1}{B_{\ell+1}}(v_{m-1}+v_{m})}{
\frac{1}{B_{\ell+1}}v_{1}+
\frac{1}{B_{\ell+1}}v_{2}+
\cdots+v_{\ell+1}+\cdots+
\frac{1}{B_{\ell+1}}v_{m}}\\
& = & \frac{v_{\ell}+v_{\ell+1}}{v_{\ell+1}}=1+\frac{v_{\ell}}{v_{\ell+1}}=1+r.
\end{eqnarray*}
This implies that for any $\varepsilon > 0$, the competitive ratio 
of {\sc greedy} cannot be less than $1+r-\varepsilon$. \BQED
%

%
%
\appendix
%
\section{Behavior of {\large GREEDY}} \label{app-g}
%
The following figure shows the behavior and the queue state of {\sc greedy} 
on the sequence $\sigma$. 
\begin{figure}[htb]
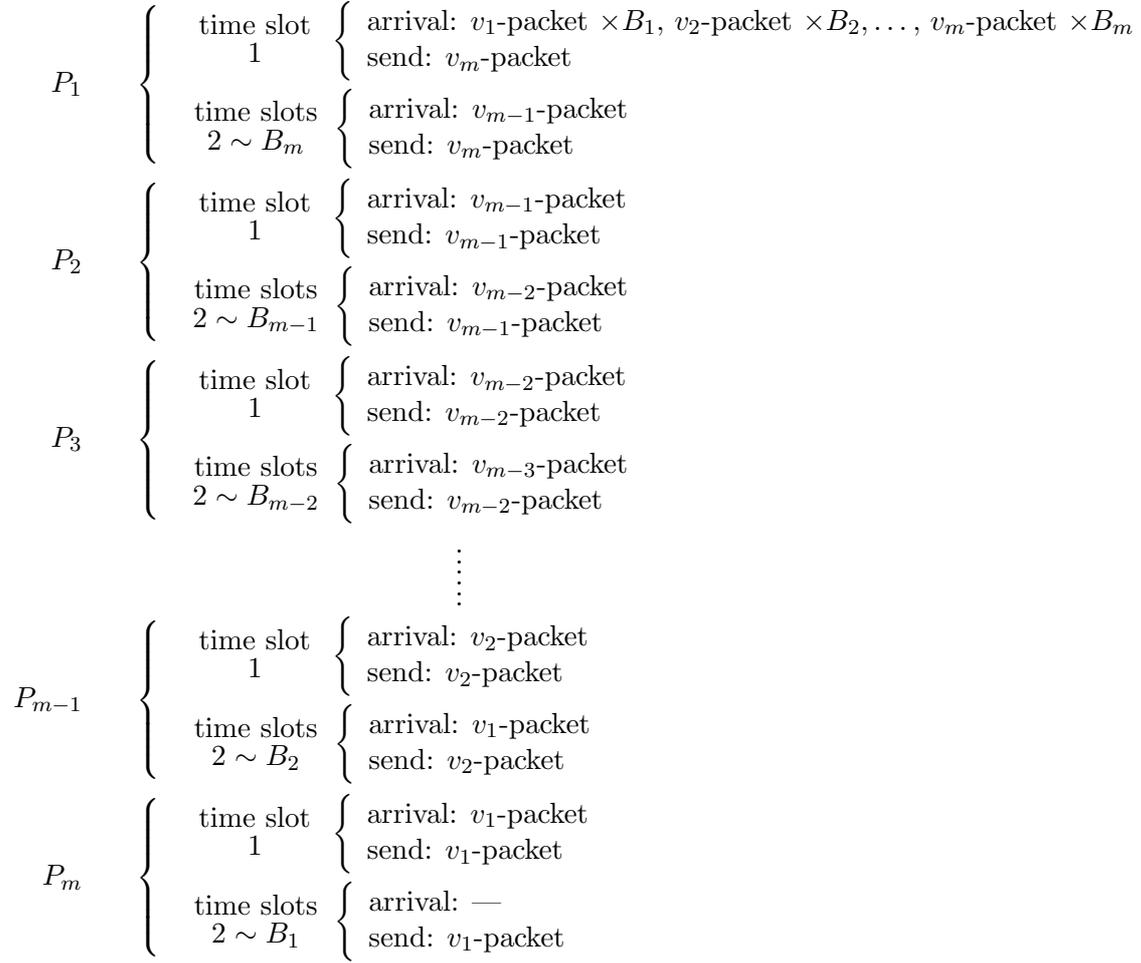

\begin{eqnarray*}
P_{1} & & \left\{
\begin{array}{l} 
\mbox{
\begin{tabular}{c}
time slot\\[-0.1cm] 1 
\end{tabular}}\left\{
\begin{array}{l}
\mbox{arrival: 
$v_{1}$-packet $\times B_{1}$, 
$v_{2}$-packet $\times B_{2},\ldots,$
$v_{m}$-packet $\times B_{m}$}\\
\mbox{send: 
$v_{m}$-packet}
\end{array} \right.\smallskip\\
\mbox{
\begin{tabular}{c}
\makebox[1.5cm]{time slots} \\[-0.1cm]  $2 \sim B_{m}$
\end{tabular}}\left\{
\begin{array}{l}
\mbox{arrival: 
$v_{m-1}$-packet}\\
\mbox{send: 
$v_{m}$-packet}
\end{array} \right.\\
\end{array} \right.\\
P_{2} & & \left\{
\begin{array}{l} 
\mbox{
\begin{tabular}{c}
time slot \\[-0.1cm] 1 
\end{tabular}}\left\{
\begin{array}{l}
\mbox{arrival: 
$v_{m-1}$-packet}\\
\mbox{send: 
$v_{m-1}$-packet}
\end{array} \right.\smallskip\\
\mbox{
\begin{tabular}{c}
\makebox[1.5cm]{time slots} \\[-0.1cm] 
\makebox[1.5cm]{$2 \sim B_{m-1}$}
\end{tabular}}\left\{
\begin{array}{l}
\mbox{arrival: 
$v_{m-2}$-packet}\\
\mbox{send: 
$v_{m-1}$-packet}
\end{array} \right.\\
\end{array} \right.\\
P_{3} & & \left\{
\begin{array}{l} 
\mbox{
\begin{tabular}{c}
time slot \\[-0.1cm] 1 
\end{tabular}}\left\{
\begin{array}{l}
\mbox{arrival: 
$v_{m-2}$-packet}\\
\mbox{send: 
$v_{m-2}$-packet}
\end{array} \right.\smallskip\\
\mbox{
\begin{tabular}{c}
\makebox[1.5cm]{time slots} \\[-0.1cm] 
\makebox[1.5cm]{$2 \sim B_{m-2}$} 
\end{tabular}}\left\{
\begin{array}{l}
\mbox{arrival: 
$v_{m-3}$-packet}\\
\mbox{send: 
$v_{m-2}$-packet}
\end{array} \right.\\
\end{array} \right.\\
& & \hspace*{4.2cm}{\vdots}\\[-0.25cm]
& & \hspace*{4.2cm}{\vdots}\\
P_{m-1} & & \left\{
\begin{array}{l} 
\mbox{
\begin{tabular}{c}
time slot \\[-0.1cm] 1 
\end{tabular}}\left\{
\begin{array}{l}
\mbox{arrival: 
$v_{2}$-packet}\\
\mbox{send: 
$v_{2}$-packet}
\end{array} \right.\smallskip\\
\mbox{
\begin{tabular}{c}
\makebox[1.5cm]{time slots} \\[-0.1cm] 
\makebox[1.5cm]{$2 \sim B_{2}$}
\end{tabular}}\left\{
\begin{array}{l}
\mbox{arrival: 
$v_{1}$-packet}\\
\mbox{send: 
$v_{2}$-packet}
\end{array} \right.\\
\end{array} \right.\\
P_{m} & & \left\{
\begin{array}{l} 
\mbox{
\begin{tabular}{c}
time slot \\[-0.1cm] 1 
\end{tabular}}\left\{
\begin{array}{l}
\mbox{arrival: 
$v_{1}$-packet}\\
\mbox{send: 
$v_{1}$-packet}
\end{array} \right.\smallskip\\
\mbox{
\begin{tabular}{c}
\makebox[1.5cm]{time slots} \\[-0.1cm] 
$2 \sim B_{1}$ 
\end{tabular}}\left\{
\begin{array}{l}
\mbox{arrival: ---}\\
\mbox{send: 
$v_{1}$-packet}
\end{array} \right.
\end{array}\right.
\end{eqnarray*}
\caption{Behavior of {\sc greedy} on $\sigma$} \label{fig-greedy}
\end{figure}
%
%
%
%
%
\newpage 
%
\section{Behavior of {\large ADV}} \label{app-a}
%
The following figure shows the behavior and the queue state of 
{\sc adv} on the sequence $\sigma$. 
\begin{figure}[htb]
\begin{eqnarray*}
P_{1} & & \left\{
\begin{array}{l} 
\mbox{
\begin{tabular}{c}
time slot \\[-0.1cm] 1 
\end{tabular}}\left\{
\begin{array}{l}
\mbox{arrival: 
$v_{1}$-packet $\times B_{1}$, 
$v_{2}$-packet $\times B_{2},\ldots,$
$v_{m}$-packet $\times B_{m}$}\\
\mbox{send: 
$v_{m-1}$-packet}
\end{array} \right.\smallskip\\
\mbox{
\begin{tabular}{c}
\makebox[1.5cm]{time slots} \\[-0.1cm] 
\makebox[1.5cm]{$2 \sim B_{m}$}
\end{tabular}}\left\{
\begin{array}{l}
\mbox{arrival: 
$v_{m-1}$-packet}\\
\mbox{send: 
$v_{m-1}$-packet}
\end{array} \right.\\
\end{array} \right.\\
P_{2} & & \left\{
\begin{array}{l} 
\mbox{
\begin{tabular}{c}
time slot \\[-0.1cm] 1 
\end{tabular}}\left\{
\begin{array}{l}
\mbox{arrival: 
$v_{m-1}$-packet}\\
\mbox{send: 
$v_{m-2}$-packet}
\end{array} \right.\smallskip\\
\mbox{
\begin{tabular}{c}
\makebox[1.5cm]{time slots} \\[-0.1cm] 
\makebox[1.5cm]{$2 \sim B_{m-1}$}  
\end{tabular}}\left\{
\begin{array}{l}
\mbox{arrival: 
$v_{m-2}$-packet}\\
\mbox{send: 
$v_{m-2}$-packet}
\end{array} \right.\\
\end{array} \right.\\
P_{3} & & \left\{
\begin{array}{l} 
\mbox{
\begin{tabular}{c}
time slot \\[-0.1cm] 1 
\end{tabular}}\left\{
\begin{array}{l}
\mbox{arrival: 
$v_{m-2}$-packet}\\
\mbox{send: 
$v_{m-3}$-packet}
\end{array} \right.\smallskip\\
\mbox{
\begin{tabular}{c}
\makebox[1.5cm]{time slots} \\[-0.1cm] 
\makebox[1.5cm]{$2 \sim B_{m-2}$} 
\end{tabular}}\left\{
\begin{array}{l}
\mbox{arrival: 
$v_{m-3}$-packet}\\
\mbox{send: 
$v_{m-3}$-packet}
\end{array} \right.\\
\end{array} \right.\\
& & \hspace*{4.45cm}{\vdots}\\[-0.05cm]
P_{m-1} & & \left\{
\begin{array}{l} 
\mbox{
\begin{tabular}{c}
time slot \\[-0.1cm] 1 
\end{tabular}}\left\{
\begin{array}{l}
\mbox{arrival: 
$v_{2}$-packet}\\
\mbox{send: 
$v_{1}$-packet}
\end{array} \right.\smallskip\\
\mbox{
\begin{tabular}{c}
\makebox[1.5cm]{time slots} \\[-0.1cm] 
\makebox[1.5cm]{$2 \sim B_{2}$}
\end{tabular}}\left\{
\begin{array}{l}
\mbox{arrival: 
$v_{1}$-packet}\\
\mbox{send: 
$v_{1}$-packet}
\end{array} \right.\\
\end{array} \right.\\
P_{m} & & \left\{
\begin{array}{l} 
\mbox{
\begin{tabular}{c}
time slot \\[-0.1cm] 1 
\end{tabular}}\left\{
\begin{array}{l}
\mbox{arrival: 
$v_{1}$-packet}\\
\mbox{send: 
$v_{1}$-packet}
\end{array} \right.\smallskip\\
\mbox{
\begin{tabular}{c}
\makebox[1.5cm]{time slots} \\[-0.1cm] 
$2\sim B_{1}$ 
\end{tabular}}\left\{
\begin{array}{l}
\mbox{arrival: ---}\\
\mbox{send: 
$v_{1}$-packet}
\end{array} \right.\\  
\end{array}\right.\\
P_{1}^{*}  & & \hspace*{0.3cm} 
\begin{array}{l}
\mbox{
\begin{tabular}{c}
\makebox[1.5cm]{time slots} \\[-0.1cm] 
$1\sim B_{2}$ 
\end{tabular}}\left\{
\begin{array}{l}
\mbox{arrival: ---}\\
\mbox{send: 
$v_{2}$-packet}
\end{array} \right.\\
\end{array} 
\\
P_{2}^{*}  & & \hspace*{0.3cm} 
\begin{array}{l}
\mbox{
\begin{tabular}{c}
\makebox[1.5cm]{time slots} \\[-0.1cm] 
$1\sim B_{3}$ 
\end{tabular}}\left\{
\begin{array}{l}
\mbox{arrival: ---}\\
\mbox{send: 
$v_{2}$-packet}
\end{array} \right.\\
\end{array}
\\
\vdots~  & & \hspace*{1.5cm} \vdots  
\hspace*{2.5cm} \vdots  \\
P_{m-1}^{*}  & & \hspace*{0.3cm} 
\begin{array}{l}
\mbox{
\begin{tabular}{c}
\makebox[1.5cm]{time slots} \\[-0.1cm] 
$1\sim B_{m}$ 
\end{tabular}}\left\{
\begin{array}{l}
\mbox{arrival: ---}\\
\mbox{send: 
$v_{2}$-packet}
\end{array} \right.\\
\end{array} 
\end{eqnarray*}
\caption{Behavior of {\sc adv} on $\sigma$}
\label{fig-adv}
\end{figure}
\end{document}